\documentclass[journal,preprint]{vgtc}                
\ifpdf
  \pdfoutput=1\relax                   
  \usepackage{graphicx}                
  \DeclareGraphicsExtensions{.pdf,.png,.jpg,.jpeg} 
\else
  \usepackage{graphicx}                
  \DeclareGraphicsExtensions{.eps}     
\fi%

\graphicspath{{figures/}{pictures/}{images/}{./}} 
\usepackage{float}
\usepackage{wrapfig}

\usepackage{color,soul}

\usepackage{microtype}                 
\PassOptionsToPackage{warn}{textcomp}  
\usepackage{textcomp}                  
\usepackage{mathptmx}                  
\usepackage{times}                     
\usepackage{cite}                      
\usepackage{tabu}                      
\usepackage{booktabs}                  

\onlineid{1028}

\vgtccategory{Research}

\title{Effect of uncertainty visualizations on myopic loss aversion and equity premium puzzle in retirement investment decisions}

\author{Ryan Wesslen, Alireza Karduni, Douglas Markant and Wenwen Dou}
\authorfooter{
\item
 Ryan Wesslen, 
 Alireza Karduni, Douglas Markant, and Wenwen Dou are with UNC-Charlotte. E-mails: \{rwesslen, akarduni,dmarkant,wdou1\}@uncc.edu.
}

\shortauthortitle{Wesslen \MakeLowercase{\textit{et al.}}: Uncertainty visualizations on myopic loss aversion}

\abstract{For many households, investing for retirement is one of the most significant decisions and is fraught with uncertainty. In a classic study in behavioral economics, Benartzi and Thaler (1999) found evidence using bar charts that investors exhibit \emph{myopic loss aversion} in retirement decisions: Investors overly focus on the potential for short-term losses, leading them to invest less in riskier assets and miss out on higher long-term returns. Recently, advances in uncertainty visualizations have shown improvements in decision-making under uncertainty in a variety of tasks. In this paper, we conduct a controlled and incentivized crowdsourced experiment replicating Benartzi and Thaler (1999) and extending it to measure the effect of different uncertainty representations on myopic loss aversion. Consistent with the original study, we find evidence of myopic loss aversion with bar charts and find that participants make better investment decisions with longer evaluation periods. We also find that common uncertainty representations such as interval plots and bar charts achieve the highest mean expected returns while other uncertainty visualizations lead to poorer long-term performance and strong effects on the equity premium. 
Qualitative feedback further suggests that different uncertainty representations lead to visual reasoning heuristics that can either mitigate or encourage a focus on potential short-term losses. We discuss implications of our results on using uncertainty visualizations for retirement decisions in practice and possible extensions for future work. 
} 


\CCScatlist{ 
 \CCScat{K.6.1}{Management of Computing and Information Systems}%
{Project and People Management}{Life Cycle};
 \CCScat{K.7.m}{The Computing Profession}{Miscellaneous}{Ethics}
}

\teaser{
  \centering
  \includegraphics[width=\linewidth]{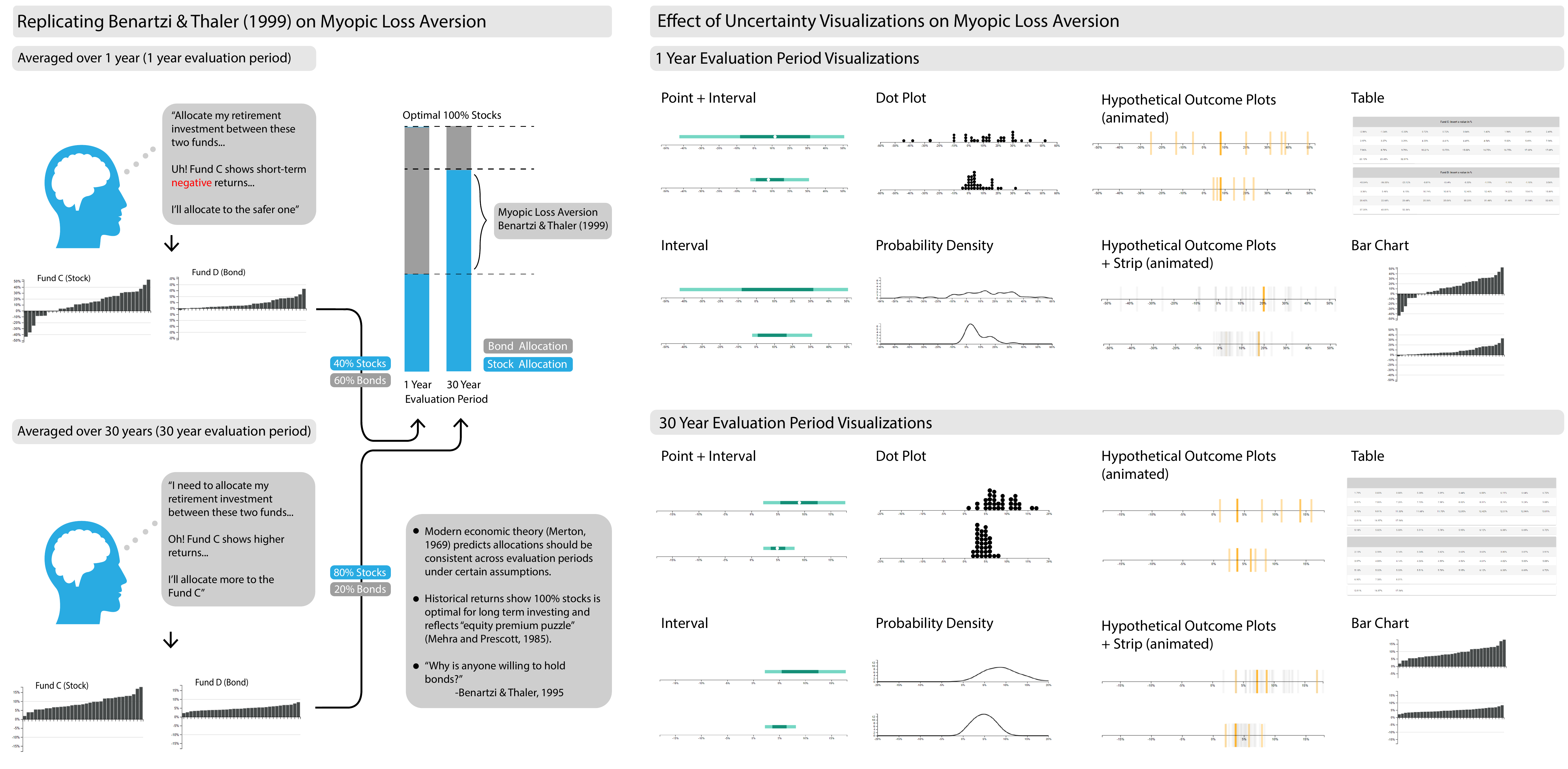}
  \caption{Overview of this study. We replicate prior work on myopic loss aversion (left) and measure the effects of different uncertainty representations (right) on myopic loss aversion in a two-round controlled and incentivized experiment.}
  \label{fig:teaser}
}

\vgtcinsertpkg

\begin{document}


\firstsection{Introduction}

\maketitle

As companies move towards choice-based retirement investment accounts like IRAs and 401(k) plans, Americans now have over \$20 trillion dollars in such accounts indicating the colossal importance of smart retirement investing.\footnote{\url{https://www.statista.com/statistics/940498/assets-retirement-plans-by-type-usa/}} A recent report by the National Institute on Retirement Security highlighted that the shift from pensions to 401(k) plans has pushed more retirement risk onto individual workers \cite{RetirementReport2020}. 
At the same time, individual investors are increasing their adoption of investing on digital platforms (e.g., online or mobile retirement account tracking). The combination of long-term retirement planning and access to short-term trends on digital investment platforms increases the importance of human-computer interaction in the design choices for web-based and mobile applications by financial companies.

Financial decision-making is central to retirement investing as typical investors make decisions about how to allocate funds across a wide range of assets that vary in risk (e.g., stocks vs. bonds). A seminal study by Mehra and Prescott \cite{mehra1985equity} found a surprising reluctance to take on risk, in that standard economic models could not account for the large historical premium for riskier investments (the ``equity premium puzzle''). Benartzi and Thaler \cite{benartzi1995myopic} theorized that individuals deviate from the predictions of neoclassical economic theory due to two factors, an oversensitivity to the possibility of losses (prospect theory), and evaluation of returns over short time periods (narrow framing), a combination they referred to as \emph{myopic loss aversion}. Benartzi and Thaler \cite{benartzi1999risk} showed that myopic loss aversion emerges when making investment decisions with bar charts of the distribution of returns. They found that investors allocated less in stocks when shown returns over a 1-year evaluation period due to aversion to short-term losses. 
Their results suggest that the method for visualizing investment performance can have a dramatic effect on individuals' willingness to take on risk. 

Recently, the field of information visualization has proposed multiple visualization techniques for including uncertainty representations of data. 
Different visual encodings can change how users perceive and interpret uncertainty, and in turn, the decisions they make.
Our study aims to measure the effect that different uncertainty visualizations/representations have on myopic loss aversion in long term (retirement) financial decision-making. The experiment is motivated by the results of Benartzi and Thaler \cite{benartzi1999risk}; we first try to replicate their findings and further expand it by testing the effect of a range of new uncertainty visualizations (e.g., quantile dotplot \cite{kay2016beyond,fernandes2018uncertainty,padilla2020uncertain} and hypothetical outcome plot \cite{hullman2015hypothetical,kale2018hypothetical,kale2020visual}) on myopic loss aversion.

More specifically, our study seeks to address the following research questions: 
\textbf{RQ1:} Do crowdsourced investors exhibit myopic loss aversion when presented investment returns that are aggregated over a range of evaluation periods (1, 5, 10, 15, 20, 25, 30 year)? RQ1 aims to first determine if we can replicate the findings of the Benartzi and Thaler \cite{benartzi1999risk} study (1 year vs. 30 year).
The visual encoding employed in evaluating RQ1 is a sorted bar chart as in \cite{benartzi1999risk}.
RQ1 extends \cite{benartzi1999risk} by evaluating the impact of intermediate evaluation periods on myopic loss aversion, providing a more fine-grained understanding of how decisions change alongside visual representations of risk.  \textbf{RQ2}: Do different uncertainty representations affect myopic loss aversion in retirement asset allocation? We included six different uncertainty representations, one bar chart visualization, and one tabular representation for RQ2. In addition to evaluating the effect of different visual encodings, RQ2 also explores how evaluation periods and visual encodings may interact. 

To address these research questions, we present findings from a two-round crowdsourced online experiment depicted in Figure 1. Round 1 is to see if we can replicate the findings of Benartzi and Thaler \cite{benartzi1999risk} on myopic loss aversion in an online setting. In round 2 participants were assigned to one of the eight different treatments with different uncertainty visualizations. 
Consistent with Benartzi and Thaler \cite{benartzi1999risk}, with bar charts we find evidence of myopic loss aversion as participants opted for much less stock allocation for 1 year than 30 year evaluation period. Similarly, we observe a positive monotonic relationship between evaluation period and stock allocation (and expected return). Interestingly, we found that the type of uncertainty representation significantly affects participants' stock allocation and their expected returns. Uncertainty representations with bar encodings led to higher stock allocation that were closer to optimal. The point + interval and the interval plot use a horizontal bar encoding of uncertainty information. Overall, the stock allocation performance with bar-encoding uncertainty representations was comparable to the bar chart employed in Benartzi and Thaler's original study \cite{benartzi1995myopic} showing the simulated returns in a sorted and non-aggregated way, the Point+Interval condition did yield stock allocation closer to the optimal in both 1 year and 30 year scenarios. 
In contrast, uncertainty representations that draw attention to risk/volatility (like hypothetical outcome plots, or HOPs) yielded lower stock allocations and tended to increase the equity premium which could amount to hundred of thousands of dollars less at retirement for an average investor. 
We did not observe significant interactions between the visual encoding and evaluation period, except for the density plot. 
Our qualitative analysis of user comments also sheds light on their visual reasoning strategies. 

Situated in the task of maximizing returns for long-term retirement planning---a common, high-stakes example of repeated decisions under uncertainty---our results demonstrate that bar-encoded uncertainty visualization with simple uncertainty representations lead to more optimal investment allocation and expected returns.
Grounded in economic theories like myopic loss aversion and the equity premium puzzle, these findings shed light on the important role of uncertainty visualization in financial decision making including retirement planning. 
%

\section{Background Work}

\subsection{Economic Theory in Long Term Investing}


Three tenets from modern economic theory provide the motivation for myopic loss aversion \cite{benartzi1995myopic,benartzi1999risk}: Samuelson's gamble \cite{samuelson1963risk}, lifetime portfolio selection \cite{merton1969lifetime,samuelson1975lifetime}, and the equity premium puzzle \cite{mehra1985equity}. During lunch one day, the noted American economist Paul Samuelson offered MIT colleagues a gamble: If the colleague guessed a coin flip correctly they would win \$200, but would lose \$100 if incorrect. 
This gamble has a positive expected value of $.5 \times \$200 + .5 \times -\$100 = \$50$.
One colleague responded that he wouldn't accept the bet once but would accept the bet 100 times. Soon after Samuelson wrote a mathematical proof \cite{samuelson1963risk} showing that such a preference was irrational and demonstrated loss aversion. For example, the faculty member turned down the bet because ``I would feel the \$100 loss more than the \$200 gain'' \cite{benartzi1995myopic}. It seems that for some individuals the fear of a loss outweighed the expected likelihood of a gain. 

\begin{figure}[t] 
\centering
\includegraphics[width=0.95\columnwidth]{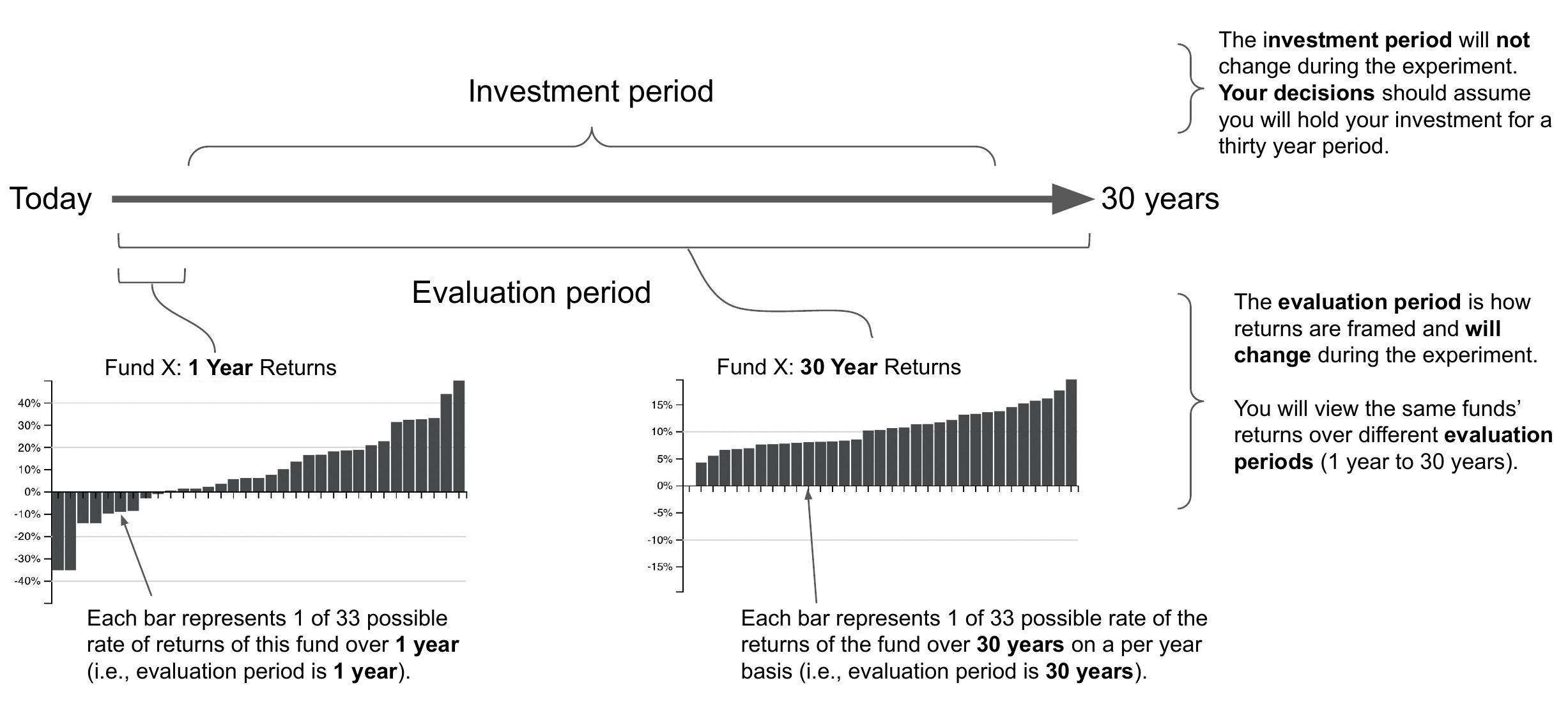}
\caption{Evaluation period versus investment period. We provided this figure to participants to distinguish between these critical concepts. Participants were instructed and incentivized to invest over a 30 year investment period for all decisions. However, following \cite{benartzi1999risk} we manipulate the evaluation period of the returns shown to participants (e.g., 1 year to 30 years). Benartzi and Thaler \cite{benartzi1999risk} argue that economic theory predicts rational investors would not differ between these two decisions but that myopic loss aversion could explain why the decisions are not consistent.}
\label{fig:period-designs}
\vspace*{-3mm}
\end{figure}

Another important tenet from economic theory of investments include theorems by Merton (1969) \cite{merton1969lifetime} and Samuelson (1975) \cite{samuelson1975lifetime} that under certain assumptions like random walk of asset prices \cite{hall1978stochastic} and constant relative risk
aversion utility functions, ``asset allocation should be independent of the time horizon of the investor'' \cite{benartzi1999risk}. This counter-intuitive notion predicts that a 35 year old and a 64 year old should choose the same allocation for retirement investing. This idea is directly connected to the differentiation between the evaluation and investment (horizon) periods, which we discuss in Figure \ref{fig:period-designs}.

Last, we draw additional motivation for our study from the perplexing financial anomaly known as the equity premium puzzle. Mehra and Prescott \cite{mehra1985equity} studied the implications of economic theory for the difference between relatively risk-free government-backed bonds (e.g., US treasury bill) and relatively more risky stocks.
They found a surprising ``equity premium'' in the form of excess returns for taking on the relatively higher risk of equity investing. 
Comparing such implications to historical returns of government bonds and stocks, Mehra and Prescott \cite{mehra1985equity} found that to account for the average 6\% equity premium, such a framework would imply an extremely risk-averse and implausible representative investor \cite{mankiw1991consumption}. Such challenges opened new behavioral explanations for the equity premium puzzle: myopic loss aversion \cite{benartzi1995myopic}.

\subsection{Investing Decisions in Behavioral Economics}

Neoclassical economic theory is grounded on the assumption of a perfectly rational decision maker \cite{landreth2002history,fox2009myth}. However, in recent decades the field of behavioral economics challenged this assumption on three fronts: unbounded rationality, unbounded willpower, and unbounded selfishness \cite{mullainathan2000behavioral}. Coined by Herbert Simon \cite{simon1955behavioral}, bounded rationality refers to the acknowledgment that individual decisions are fundamentally constrained by factors like scale, time, and cognitive ability. 
This work led to the study of heuristics, or mental shortcuts, which people use to find solutions for decisions otherwise thought to be intractable. 
Tversky and Kahneman \cite{tversky1974judgment} argued that such heuristics can lead to systematic errors in decision-making, or cognitive biases. They extended such work with prospect theory, a descriptive theory of decision-making  under uncertainty \cite{kahneman1979prospect}.
A main assumption of prospect theory is that utility is derived not from objective metrics of value, but instead as gains and losses relative to some reference point.
Prospect theory captures people's inherent risk aversion or ``loss aversion'', the notion that individuals are affected more by a loss than a gain of the same magnitude.  
A key implication is that changes in the \emph{framing} of a choice in terms of gains or losses---despite no differences in the underlying economic problem---can exert a strong influence on decisions \cite{tversky1986RationalChoiceFraming}.

Benartzi and Thaler \cite{benartzi1995myopic} proposed that a narrow framing of investment decisions focused on short-term outcomes rather than long-term (aggregated) returns, in combination with loss aversion, provides a solution to the equity premium puzzle.
Evaluating returns over a short time window highlights the possibility of losses and leads to more risk aversion, even in the context of investment decisions with long time horizons.
Using prospect theory, Benartzi and Thaler \cite{benartzi1995myopic} found that the equity premium observed by Mehra and Prescott was consistent with an evaluation period of 1 year.

As Benartzi and Thaler noted, while loss aversion might be considered a matter of individual preference (or a ``fact of life''), the evaluation period is often  dictated by the environment or ``choice architecture'' \cite{thaler2008nudge}.
Accordingly, subsequent work has shown that myopic loss aversion is reduced when investors are provided aggregated returns over longer evaluation periods \cite{benartzi1999risk,thaler1997effect} or when they have less frequent opportunities to change their allocations \cite{hardin2012myopic}.
These effects exemplify a common theme in behavioral economics: Subtle features of the choice architecture can “nudge” people toward behaviors that are consistent with a policy objective \cite{thaler2008nudge}, such as maximizing retirement savings through the use of default contributions \cite{thaler2004save}. 
For instance, in a recent study using a simulated retirement investment task, Camilleri et al. \cite{camilleri2019NudgesSignpostsEffect} found that presenting dynamic risk information, such that the evaluation period was aligned with the time left until retirement, encouraged reliance on a “smart default” plan which invested more in riskier (higher growth) investments early on. 
These results underscore the importance of interaction design in investment platforms.
Providing frequent updates on short-term changes in fund performance and making it easy to reallocate may increase engagement, but these same factors likely amplify the harmful effects of myopic loss aversion on long-term returns \cite{looney2009DecisionSupportRetirement}.


\subsection{Visualization in financial decisions and uncertainty visualization}

Research in data visualization and human-computer interaction has examined the role of data visualizations in improving financial decision-making \cite{savikhin2008applied,rudolph2009finvis,savikhin2011experimental,ko2016survey,savikhin2013application}. Gunaratne and Nov \cite{gunaratne2015informing} developed a user interface for retirement decision-making based on endowment effect and loss aversion.
They showed that highlighting the long-term implications of users' decisions could lead to adjustment of their investment allocations and reduction of loss aversion. Related HCI research on the gamification of uncertainty decisions has shown that too much uncertainty information can lead to unnecessary risk-taking \cite{greis2016decision}. Other research has considered the role of portfolio allocation or financial literacy. Rudolph \textit{et al.} \cite{rudolph2009finvis} designed a financial planning study using a simple visual analytics system (FinVis) to aid in portfolio allocation. Using a simple table as a control group, they found university students made more optimal allocation decisions using FinVis as compared to the control (table of returns and standard deviation). Lusardi \textit{et al.} provides visual analytics tools for financial literacy \cite{lusardi2017visual}. Additional research has also considered the intersection of visualization and financial decisions in relation to cognitive biases \cite{zhang2015designing} and risk premium \cite{yue2019sportfolio}. But one gap of this research on financial decision-making in visualization research has been the connection with research on visualizing uncertainty in which different techniques for encoding uncertainty such as Hypothetical Outcome Plots \cite{hullman2015hypothetical} or Quantile Dot Plots \cite{fernandes2018uncertainty} which have different effects on users' belief-updating and decision outcomes \cite{kay2016beyond,fernandes2018uncertainty,kale2018hypothetical,kale2020visual,karduni2020bayesian}. Kay \textit{et al.}  \cite{kay2016ish} noted that uncertainty can be intrinsic or extrinsic to the representation which we found informative for our study.  
For a comprehensive survey on uncertainty visualizations, we recommend Padilla, Hullman, and Kay \cite{padilla2020uncertain}. 

\section{Research questions and hypotheses}

The core research question of this work follows Benartzi and Thaler \cite{benartzi1999risk}: ``How do investors think about investment decisions over long horizons, and how do their choices depend on the way in which risk and return data are presented?'' From this question, we derive two research questions.


\textbf{RQ1}: Replicating Benartzi and Thaler (1999), do crowdsourced investors exhibit myopic loss aversion when presented with a 1 year versus a 30 year evaluation period? More broadly, what is the effect of different evaluation periods on myopic loss aversion?

\textbf{Design and Hypothesis}: MTurk participants who own at least one financial asset participate in a within-subject asset allocation decision between stock and bond returns (both names masked) over seven evaluation periods (i.e. 1, 5, 10, 15, 20, 25, 30 year) for a fixed 30 year investment period. 
Myopic loss aversion \cite{benartzi1995myopic,benartzi1999risk} predicts that individuals are more risk averse for shorter evaluation periods (e.g., 1 year) than longer periods, resulting in a higher asset allocation to less risky assets (e.g., bonds) than is optimal for long planning horizons like retirement.

\textbf{RQ2}: Does visualization with uncertainty representation affect myopic loss aversion (i.e., retirement asset allocation) and do uncertainty representations interact with evaluation periods?

\textbf{Design and Hypothesis}: Consistent with past designs for uncertainty visualizations \cite{kay2016ish,fernandes2018uncertainty,padilla2020uncertain}, we design a mixed experiment with between-subjects (different uncertainty visualization) and repeated measures within-subjects (seven evaluation periods). We expect that visualizations with intrinsic uncertainty representations like frequency framing (e.g., dotplot) and animated plots (e.g., hypothetical outcome plots) will result in better returns and less myopic loss aversion.



\section{Methods}

We designed a pre-registered\footnote{\url{https://aspredicted.org/6zj4b.pdf}} experiment to test how uncertainty visualizations impact allocation decisions for simulated long term (30 year) retirement investments.
Following Benartzi and Thaler \cite{benartzi1999risk}, we frame individuals' decisions as dividing an investment between two assets. 
The names of the assets are masked but they correspond to standard benchmarks for bonds (10 year United States Treasury) and stocks (S\&P 500).\footnote{Data is from Aswath Damodaran and available at  \url{http://pages.stern.nyu.edu/~adamodar/New_Home_Page/datafile/histretSP.html}.}
We compared the bar chart visualization from Benartzi and Thaler \cite{benartzi1999risk} with a range of alternative uncertainty representations.

\subsection{Investment task and experiment design}

\begin{figure}[t] 
\centering
\vspace*{-3mm}
\includegraphics[width=0.95\columnwidth]{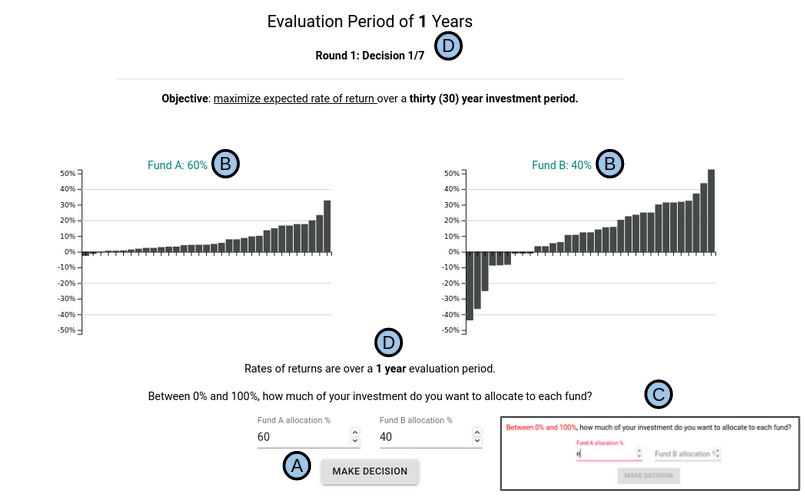}
\caption{A depiction of the experiment interface. This example shows the round 1 (bar chart) and 1 year evaluation period decision. The user inputs their allocation (A) for each evaluation period (D) that updates chart titles (B) and input is controlled for invalid responses (C).}
\label{fig:interface}
\vspace*{-3mm}
\end{figure}

The experiment included two rounds. The first round was a within-subjects manipulation of evaluation period using the bar chart visualization from \cite{thaler2015misbehaving} (see example in Figure~\ref{fig:interface}). 
Because we aimed to replicate \cite{benartzi1999risk}, the first two decisions were based on 1 year or 30 year evaluation periods presented in random order. 
The remaining decisions involved five different evaluation periods in a random order (5, 10, 15, 20, and 25 years). 
In the second round, each user was randomly assigned to one of eight uncertainty visualization conditions.
The presentation order followed the same scheme as in Round 1 (1 and 30 years in the first two trials, followed by the remaining evaluation periods).
Participants were not provided immediate feedback about their decisions (i.e., simulated returns based on their allocations) in order to avoid any learning effects. To aid participant understanding of the task, we provided simple task instructions as well as Figure \ref{fig:period-designs} to distinguish between the evaluation and investment periods. We then asked three questions following round 1 instructions: what is the hypothetical investment period (30 years); what is the basic task (allocate retirement investment); what leads to higher incentives (higher simulated returns). Participants could not move forward if incorrect but could modify their answers until they provided the correct answers.

We designed a react.js custom web interface that used D3 for visualizations, node.js for the server, mongoDB for the database and the app was deployed on heroku.\footnote{The experiment is available at \url{https://retirement-study-1.herokuapp.com/} and the code is \url{https://github.com/wesslen/financial-decisions}.} Figure \ref{fig:interface} provides a screenshot of the investment task interface for an evaluation period of 1 year.
The charts show the distribution of simulated returns for each fund over the evaluation period.
The left-right position of the two assets (bonds vs. stocks) was randomized on each trial.
The participant inputs an allocation in either of two boxes (A) that are reactive (i.e., sum of the two boxes always equal 100\%). As the user enters valid value (0 to 100 in 1 increments), the ``Make Decision'' button becomes active and the user can proceed to the next allocation decision. 
To ensure understanding of their decision, each chart's title updates real time based on participant's decision (B). 
If the user enters invalid input (e.g., ``e''), ``Between 0\% and 100\%'' is highlighted in red and ``Make Decision'' becomes inactive, preventing the user from moving to the next decision until a valid response is provided. 
For each decision, the interface updates the evaluation period and emphasize it in two places (D). 

There are three intentional deviations from the original experiment by Benartzi and Thaler \cite{benartzi1999risk}. First, we provide incentives to align to performance. This is critical to ensure participants have a vested interest in performing this task. This change is especially important given the use of crowd sourced workers. Second, for round 1, we use the bar chart design from \cite{thaler2015misbehaving} which lays out each fund in its own bar chart horizontally rather than the original design which was one grouped bar chart with each fund being a different bar. We did this for design preferences. Third, the original study used 34 bars for the 1 year return and 50 bars for the 30 year return. For simplicity, we used 33 draws (e.g., bars, dots) for easy mental computation that each draw is approximately 3\% of the data. We also controlled 33 draws across visualizations and evaluation periods for consistency. 

\begin{figure}[t] 
\centering
\vspace*{-3mm}
\includegraphics[width=0.95\columnwidth]{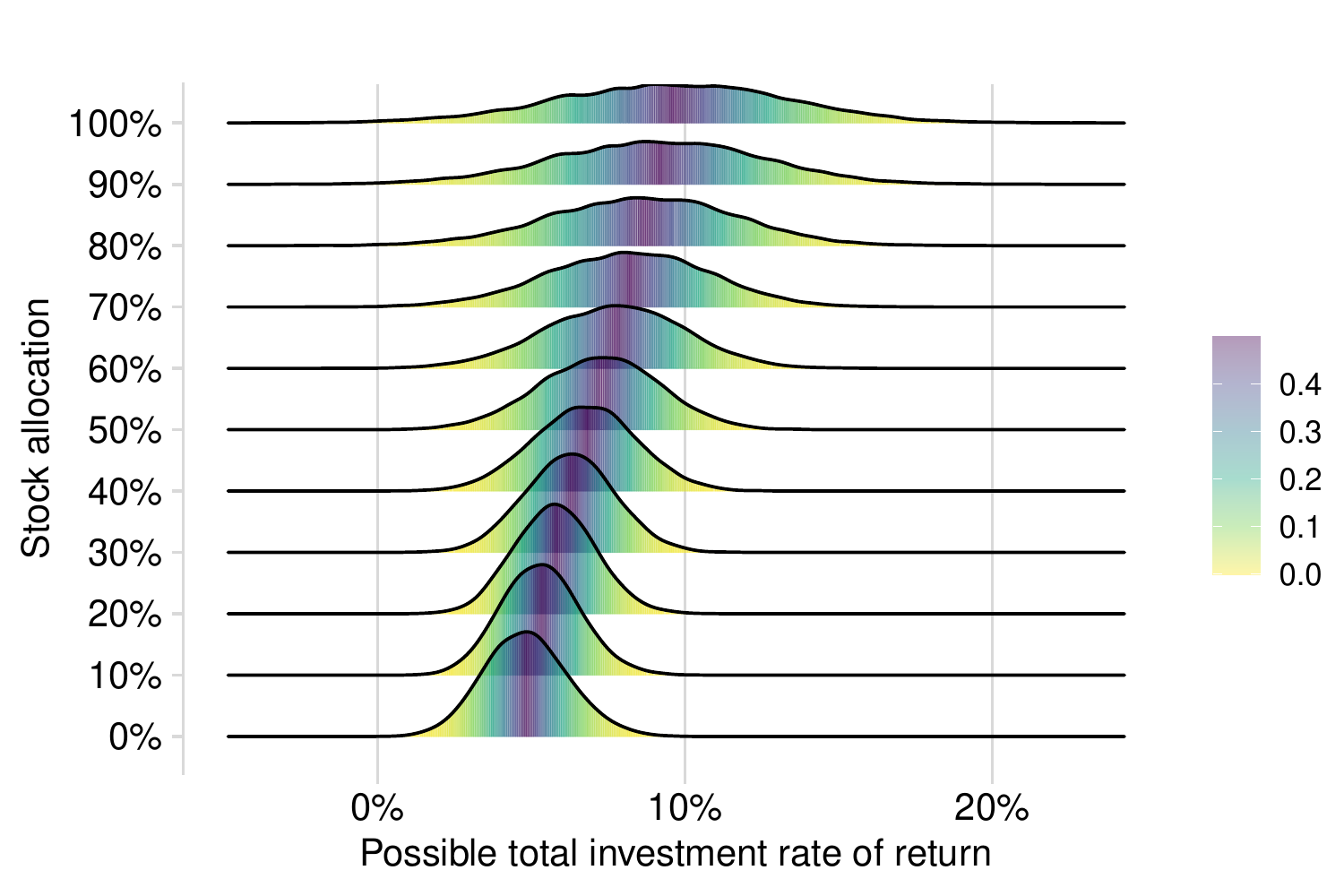}
\caption{Historical simulation of returns for different stock allocation (S\&P500) decisions over 30 year investment period in 10\% increments. Bond allocation (10 year US Treasury) is 1 minus stock allocation. We use an empirical cumulative density function with viridis palette to indicate density. 100\% stocks is the optimal allocation for maximizing expected returns and is consistent with the equity premium puzzle \cite{mehra1985equity}.}
\label{fig:sim-returns}
\vspace*{-3mm}
\end{figure}

\subsubsection{Simulating expected returns}

Following \cite{benartzi1999risk}, we showed participants data generated from a historical simulation using bootstrapped sampling with replacement. To calculate average and annualized returns by evaluation period \textit{N}, we used the geometric mean for each sample \textit{i}: 

\begin{equation}
\textrm{Geometric Mean} =\left(\prod _{i=1}^{N}(1+Returns_{i})\right)^{\frac {1}{n}}-1
\end{equation}

Participants decided the allocation between stocks and bonds for a simulated 30 year investment for retirement.
Incentives were established by running a 10,000 bootstrap with replacement sample returns with a 30 year investment period. 
We then calculated what the expected return would be for 101 different possible stock allocations (0 to 100 in integer increments), with the remainder allocated to bonds. 
When participants made a decision, we determined their incentive by randomly selecting one of the conditional returns given their chosen allocation and the 30-year investment period. 

We used the same two asset simulation across 30 years to derive the distribution of portfolio outcomes and calculated the average return for each of the possible allocation combinations. Figure \ref{fig:sim-returns} shows the distribution of simulated returns for stock allocations in increments of 10\%. The color represents density with mean/median corresponding to the darker areas. Aligned with the equity premium puzzle \cite{mehra1985equity}, 100\% stock allocation has the highest expected return and is the optimal decision if maximizing the expected return. 
The expected return is also a monotonic function of the stock allocation decision.

One drawback of using stock allocation as a dependent variable is that it does not reflect the participants' expected returns and how that decision compares to the optimal strategy of choosing 100\% stocks. 
To address this drawback, we follow Fernandes \textit{et al.} \cite{fernandes2018uncertainty} and convert stock allocations into the ratio of the expected return relative to the optimal strategy (i.e., expected return for 100\% stocks). 
We then use this ratio of expected return to optimal expected return as our decision metric and dependent variable in our regression analyses.

\subsubsection{Uncertainty representations}

In the second round of the task we evaluated the effect of different uncertainty representations on retirement financial decision making. 
Our choices of uncertainty representations were grounded in prior work  \cite{kay2016ish, jung2015, Wunderlich2017} such that the representations were designed to be ``glanceable'' and to allow ``quick in-the-moment decisions to be made'' without significant training. The selected uncertainty representations cover multiple categories discussed in  \cite{kay2016ish}, including uncertainty representations that are intrinsic and with extrinsic annotation;  uncertainty representations as continuous and discrete outcomes. Although subsets of the selected uncertainty representations have been evaluated/compared in previous work with different decision making tasks, including bus arrival [7, 55] and job growth [21], we think it’s worthwhile to include a comprehensive set of uncertainty representations in the new task of making retirement decisions.
The resulting uncertainty representations cover a design space characterized by frequency framing \cite{kay2016beyond,fernandes2018uncertainty}, point-interval \cite{kale2020visual}, animation \cite{hullman2015hypothetical,kale2020visual}, and controls (round 1 bar chart and table). To ensure consistency across the visualizations, all of the discrete plots use the same 33 returns from bootstrap sampling with replacement. We discuss the rationale of our decision to include each of the visualizations below as our eight treatments.

\textbf{Table}: As a control treatment, we provided a table of the returns to compare uncertainty visualizations to a treatment with no data visualization. This treatment enables us to measure the marginal effects of data visualizations over the raw data. The tables provides the data in ascending order from lowest (top left) to highest (bottom) returns.

\begin{figure}[H] 
\centering
\vspace*{-3mm}
\includegraphics[width=0.8\linewidth]{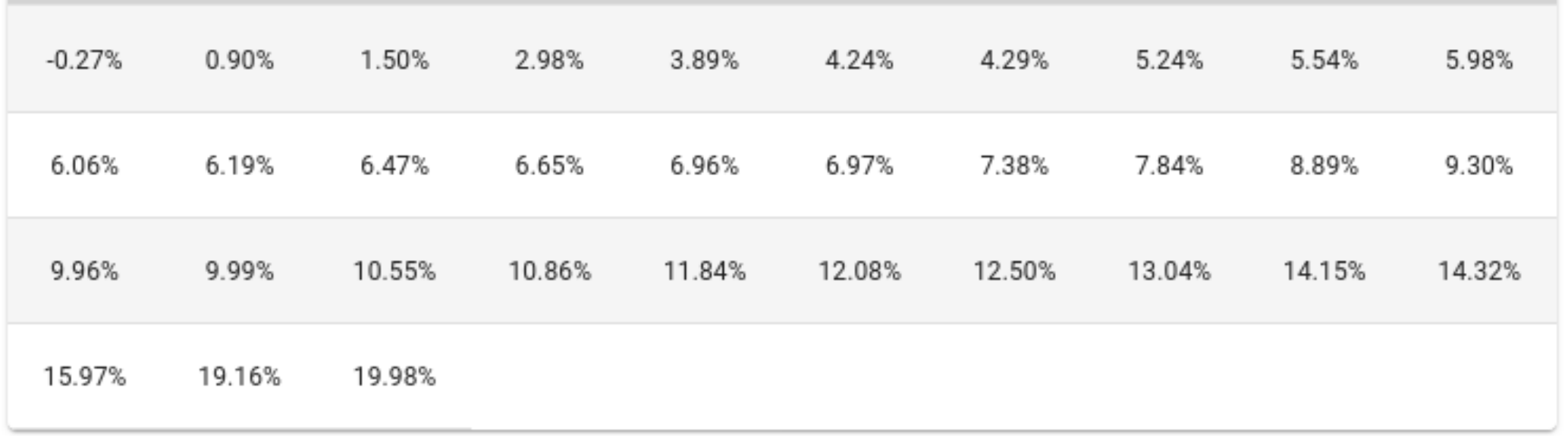}
\label{fig:table-example}
\vspace*{-3mm}
\end{figure}

\textbf{Bar chart}: We repeated the same visualization from round 1 (see Figure~\ref{fig:interface}) to enable a between-subjects comparison of the bar chart versus other uncertainty visualizations.


\textbf{Point-Interval}: Recent research in uncertainty visualizations on effect size judgments has shown that providing means to uncertainty visualizations has possible biasing effects \cite{kale2020visual}. To test for such an effect in our task, we developed the point-interval condition that showed participants intervals (bars) that represent a range containing 66\% (dark green) and 95\% (light green) of the possible (bootstrapped) outcomes as well as a point estimate of the mean.  

\begin{figure}[H] 
\centering
\vspace*{-3mm}
\includegraphics[width=0.8\linewidth]{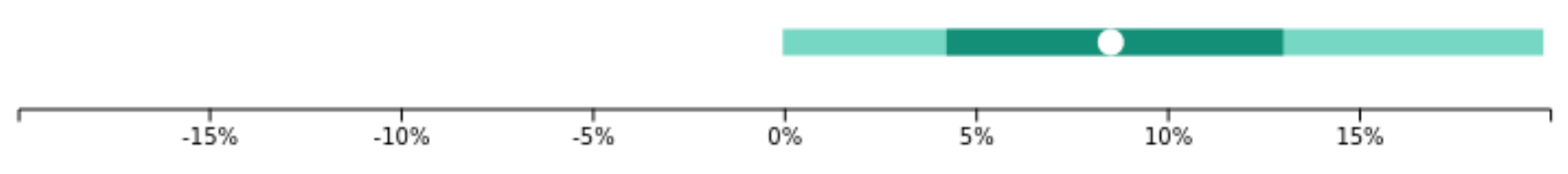}
\label{fig:point-example}
\vspace*{-3mm}
\end{figure}

\textbf{Interval}: Similar to the point-interval, we provided another condition with the same intervals (66\% and 95\%) but no point (mean) estimate. Without a separate encoding for the mean, users could only mentally estimate the mean as the midpoint of the intervals. 

\begin{figure}[H] 
\centering
\vspace*{-3mm}
\includegraphics[width=0.8\linewidth]{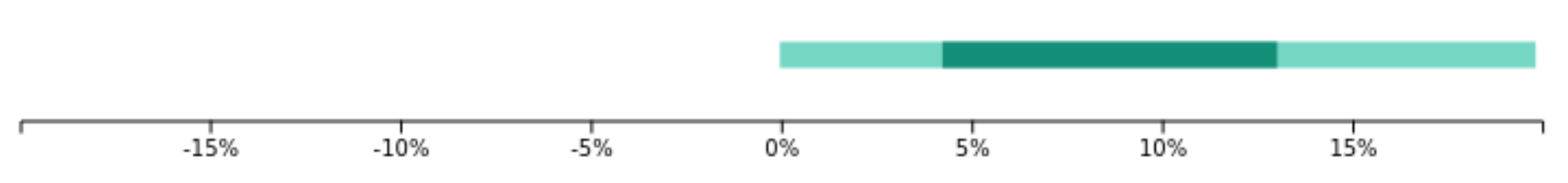}
\label{fig:interval-example}
\vspace*{-3mm}
\end{figure}

\textbf{Dot plot}: Past research \cite{kay2016ish,fernandes2018uncertainty,kale2020visual,padilla2020uncertain} has shown that frequency framing \cite{gigerenzer1996psychology} can improve the understanding of probabilities better than other representations in a variety of tasks. We provide a Wilkinson dot plot \cite{wilkinson1999dot} to display each dot as one of the 33 possible sampled returns.\footnote{While similar to a quantile dot plot\cite{kay2016ish,fernandes2018uncertainty}, we label it a Wilkinson plot instead of a quantile dot plot as it was generated from a discrete (historical) distribution, not from a continuous distribution.}
 
 \begin{figure}[H] 
 \centering
\includegraphics[width=0.8\linewidth]{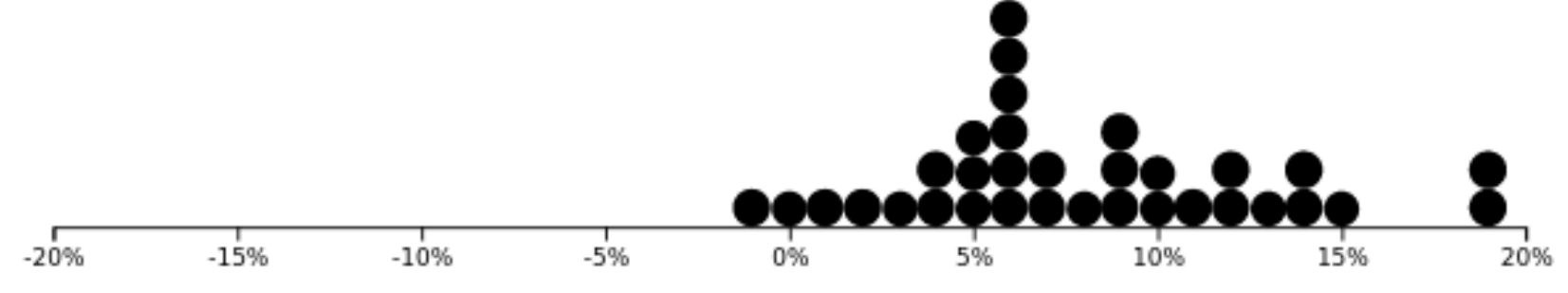}
\label{fig:dots-example}
\vspace*{-3mm}
\end{figure} 
 
\textbf{Probability density}: A popular data visualization to represent uncertainty on distribution is the (probability) density plot. The purpose of a density is to visualize an underlying data distribution through an approximate continuous curve. This enables a smooth representation that is common in probability distributions and used in spatial plots \cite{li2017streammap}.

\begin{figure}[H] 
\centering
\vspace*{-3mm}
\includegraphics[width=0.8\linewidth]{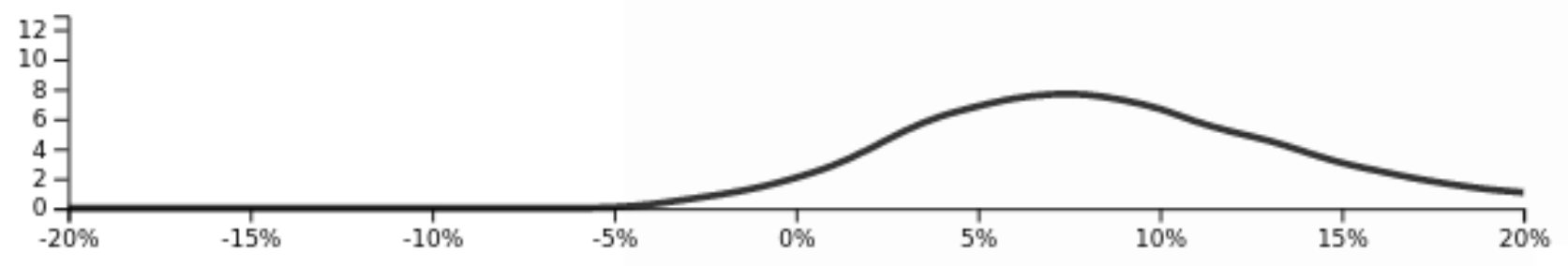}
\label{fig:density-example}
\vspace*{-3mm}
\end{figure}

\textbf{HOP}: While static visualizations like error bars are the most common uncertainty visualization, hypothetical outcome plots (HOPs) are designed to focus on the user experiencing uncertainty information through animated draws. HOPs have been studied in a variety of applications including identifying trends \cite{kale2018hypothetical}, effect size judgments \cite{kale2020visual}, and correlation judgment \cite{karduni2020bayesian}.

\begin{figure}[H] 
\centering
\vspace*{-3mm}
\includegraphics[width=0.8\linewidth]{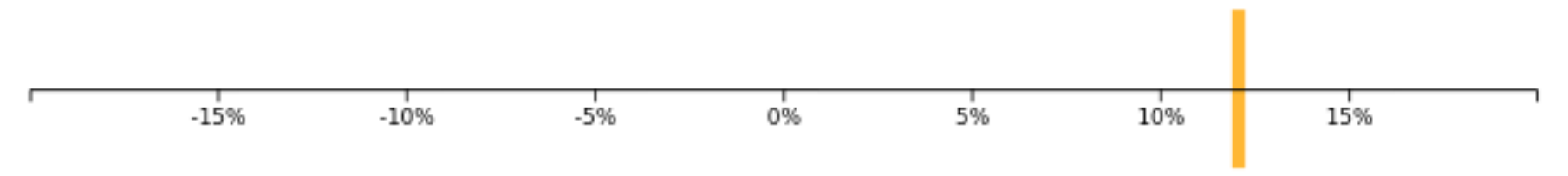}
\label{fig:hops-example}
\vspace*{-3mm}
\end{figure}

\textbf{HOP + Strip}: We also provide a hybrid HOP that combines a static distribution (strip plot) overlaid with the HOP. The motivation of this plot is that we expect individuals will perform better in this condition than the HOP only as it reduces the cognitive load to storing into short term memory the sampled distribution and enables the user to focus attention on the draws relative to the sampled distribution (strip plot).

\begin{figure}[H] 
\centering
\vspace*{-3mm}
\includegraphics[width=0.8\linewidth]{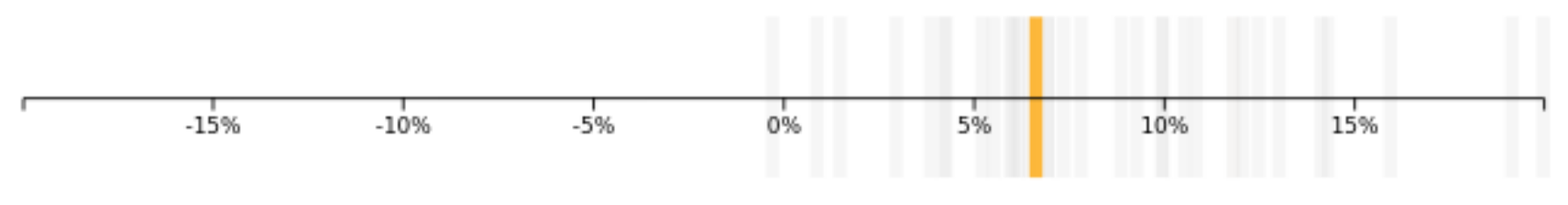}
\label{fig:hopsdist-example}
\end{figure}

\subsection{Participants}

We recruited participants through Amazon Mechanical Turk. 
Given our focus on financial decision-making, we targeted MTurk workers who report owning at least one of four financial products: stocks, bonds, mutual funds, or electronic traded funds (ETFs). To ensure a high quality of performance, participants were either a MTurk Master ($n=42$) or a non-Master with a HIT acceptance rate of 97\% or better ($n=179$). Following \cite{benartzi1999risk}, we targeted a sample size of 25 participants per condition (target 200 total across eight conditions). 
221 participants completed the study as we expected exclusions.
After applying our pre-registration exclusions, we ended with 198 total participants. 
The average total compensation (with bonus) was \$2.46 and participants took on average 14.6 minutes to complete the study. The average age was 37.4 years and 27\% identified as female.
Out of the entire sample, 67.6\% reported owning a retirement investment account.

\begin{figure}[t] 
\centering
\includegraphics[width=0.95\columnwidth]{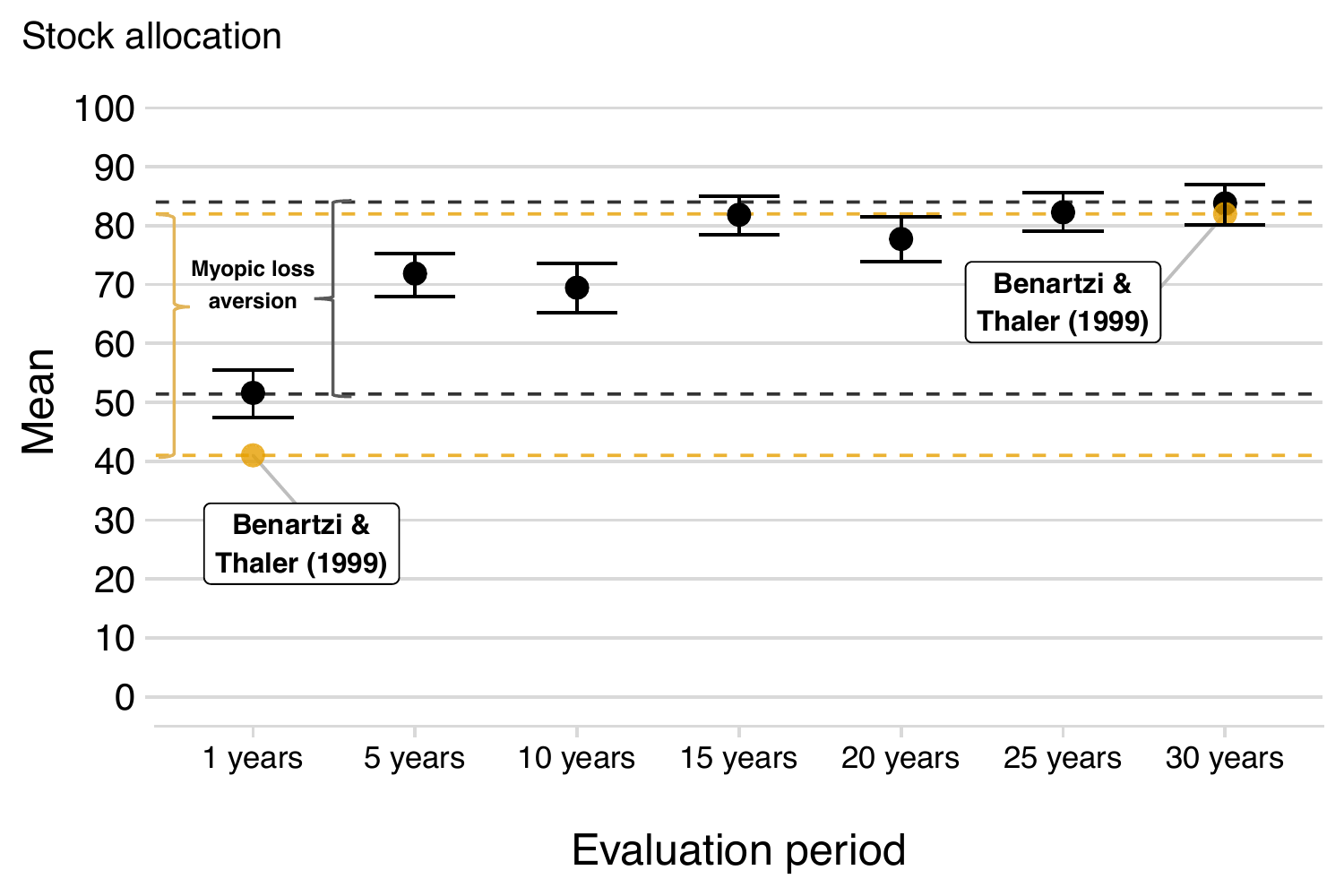}
\caption{Round 1 mean stock allocation and bootstrapped 95\% confidence intervals ($n = 198$) by evaluation period by participant. The orange points are the original values from Benartzi and Thaler \cite{benartzi1999risk}. Dotted lines are means for 1 and 30 year evaluation periods and the arrows indicate the allocation difference which we measure as myopic loss aversion.}
\label{fig:boot-rq1}
\vspace*{-3mm}
\end{figure}

\begin{figure*}[!t] 
\centering
\includegraphics[width=0.8\linewidth]{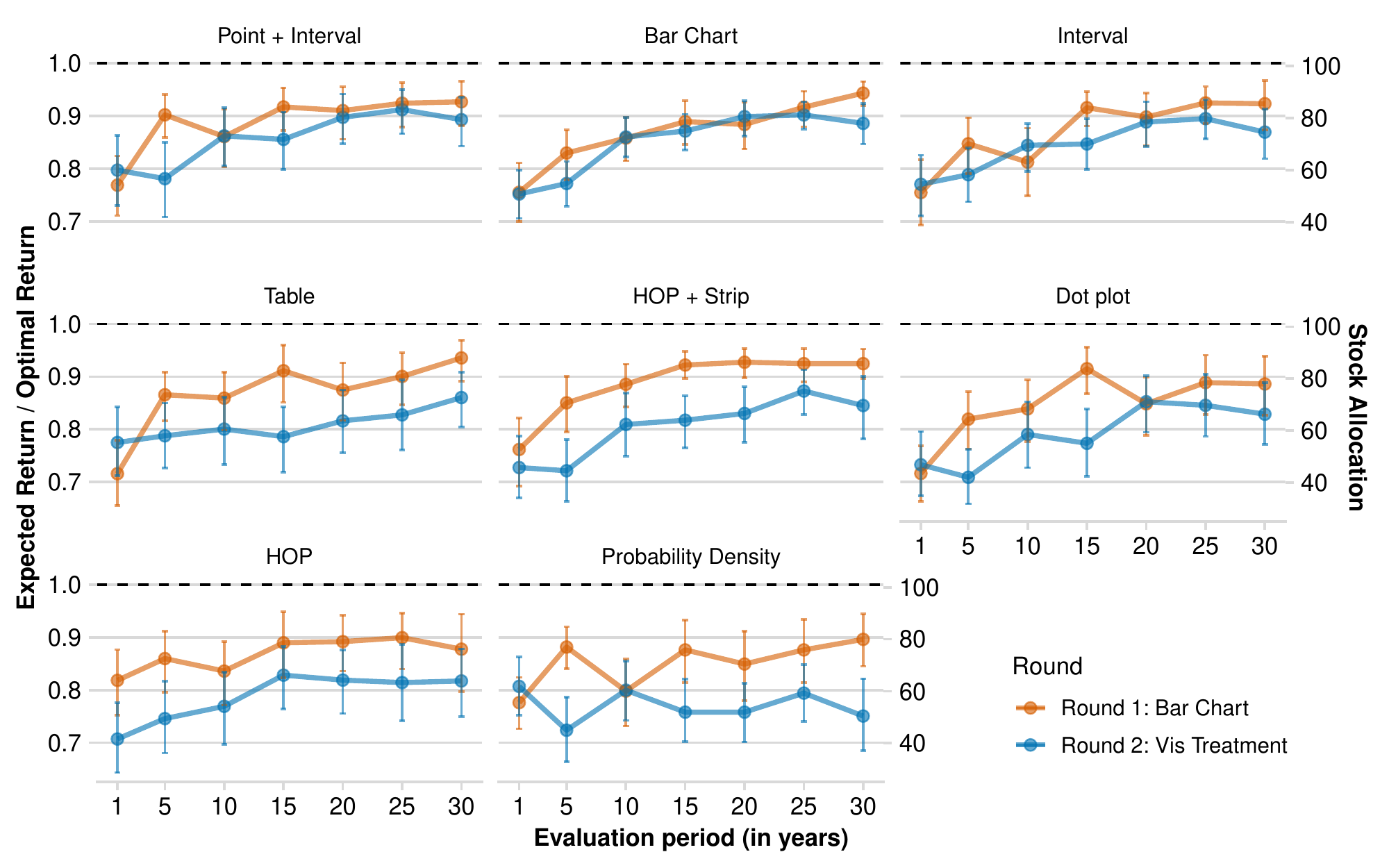}
\caption{Mean and bootstrapped 95\% confidence intervals for participants' expected / optimal return (left axis) and stock allocation (right axis). We provide results for both round 1 (bar chart only) and round 2 (visualization treatment). The dotted line indicates the optimal strategy (100\% stocks).}
\label{fig:bootstraped-values}
\vspace*{-3mm}
\end{figure*}

\subsection{Procedure}

In addition to the main task, the experiment also included:


\textbf{Post-questionnaire}: We required participants to answer six closed ended demographics questions to measure sex, age, education, ownership of financial asset (e.g., stock, bond, ETF, mutual fund), ownership of retirement investment account (e.g., 401k, IRA, Roth IRA), and satisfaction with the study. We also solicited (optional) open ended user feedback on the study.

\textbf{Payment}: All participants who completed the study received at least \$1.00 (base) + a bonus of up to \$3.50 based on simulated performance (up to \$0.25 per 14 trial). 
A participant's bonus from a trial was based on the quintile of simulated performance of their allocation, with the bonus increasing by \$0.05 at each quintile (i.e., lowest quintile: \$0.05; highest quintile: \$0.25). 


\textbf{Attention/Learning Checks}: First, we included a CAPTCHA to screen for bots and as an attention check. Second, as mentioned in Section 4.1, participants were required to complete three comprehension questions. Third, to check for knowledge of the bar chart interpretation, in round 1 participants were asked to hover their mouse over the largest bar and provide the value of that bar. Participants could not proceed without the correct answer (rounded to nearest whole number). 

\textbf{User strategy feedback}: Recent work has shown users having challenges with uncertainty visualizations through suboptimal strategies or switching strategies \cite{kale2020visual}. Following \cite{kale2020visual}, we asked participants the following qualitative question to elicit feedback on user strategies after each round: ``How did you use the charts to complete the task? Please do your best to describe what sorts of visual properties you looked for and how you used them?''

\subsection{Analysis approach}

Following Fernandes \textit{et al.} \cite{fernandes2018uncertainty}, we use a mixed effects Bayesian beta regression. We used a beta regression given that the ratio of the expected return to the optimal expected return are values between 0 and 1. We included the participant ID as a random effect given that decisions are repeated by participant. For fixed effects, we considered both evaluation period and visualization treatment and interaction between evaluation period and treatment.\footnote{Per our pre-registration, we considered two variants with and without the interaction and decided to include the interaction due to a lower model AIC.} 
For our priors, we considered both non-informative priors and priors from Fernandes \textit{et al.} \cite{fernandes2018uncertainty} but there were no substantive differences. 
For model fitting and visualization we used R packages \textbf{tidyverse} \cite{wickham2019welcome}, \textbf{brms} \cite{burkner2017brms} and \textbf{tidybayes} \cite{kay2019tidybayes}.



To aid in the interpretation of our results, we convert the dependent variable into the expected investment value for a hypothetical investor at retirement. Consistent with our average age (37), we assume someone who will retire in 30 years which typically occurs in the United States between 65 and 67. Also we assume an initial investment balance (\$50,000) similar to the average retirement savings for that age group (\$48,710). For simplicity, we assume no role of taxes, no need for liquidity (e.g., cash out), zero discount rate or no inflation, and no additional contributions or withdrawals. 



\begin{figure*}[!t] 
\centering
\includegraphics[width=0.95\linewidth]{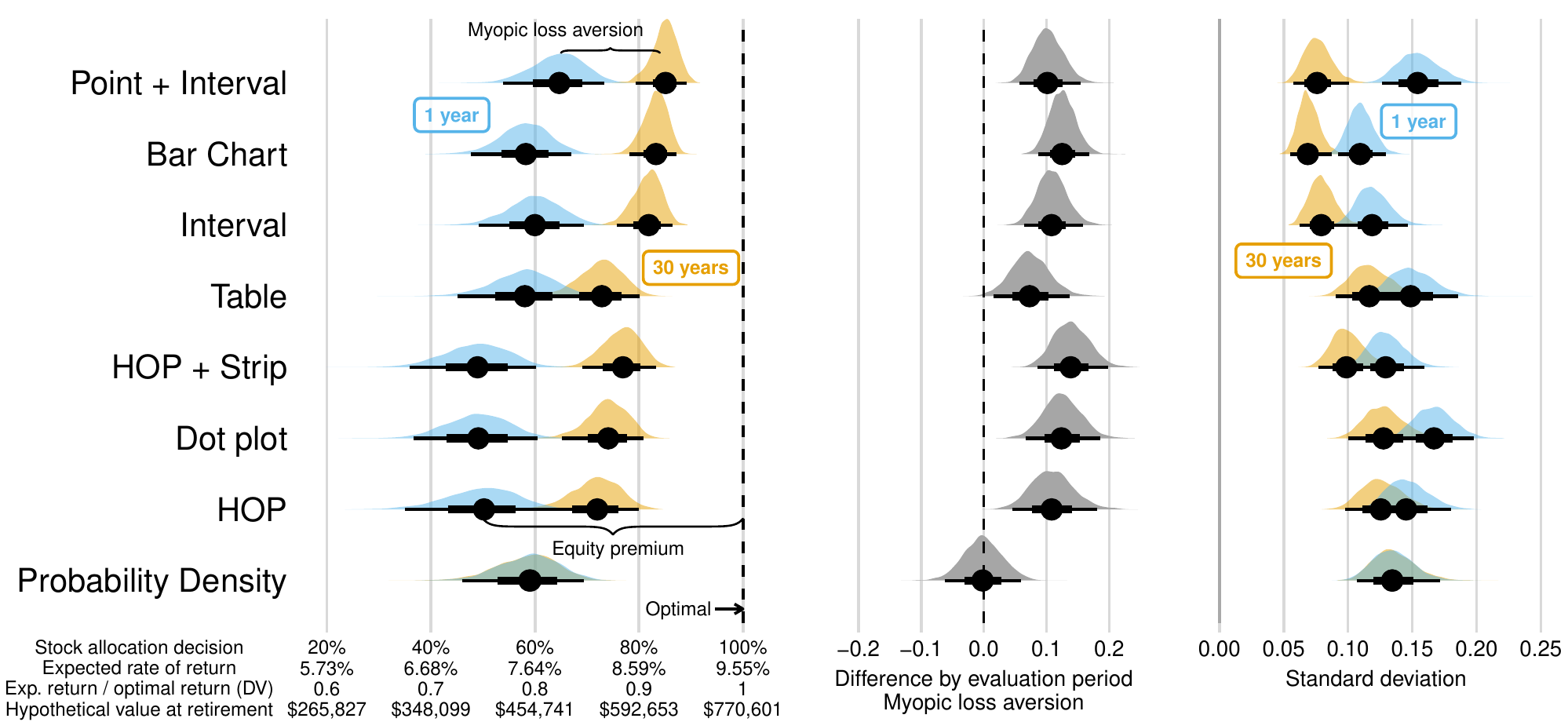}
\caption{Round 2 posterior mean (mu), mean differences, and standard deviations by 1 year (blue) and 30 year (orange) evaluation periods by treatment. We provide multiple conversions of the DV (expected return / optimal return) including the expected return, the stock allocation, and the retirement balance for a hypothetical 37 year old with a \$50,000 initial investment (subject to other assumptions).}
\label{fig:eval_end_plots}
\vspace*{-3mm}
\end{figure*}

\begin{figure*}[t] 
\centering
\vspace*{-3mm}
\includegraphics[width=0.80\paperwidth]{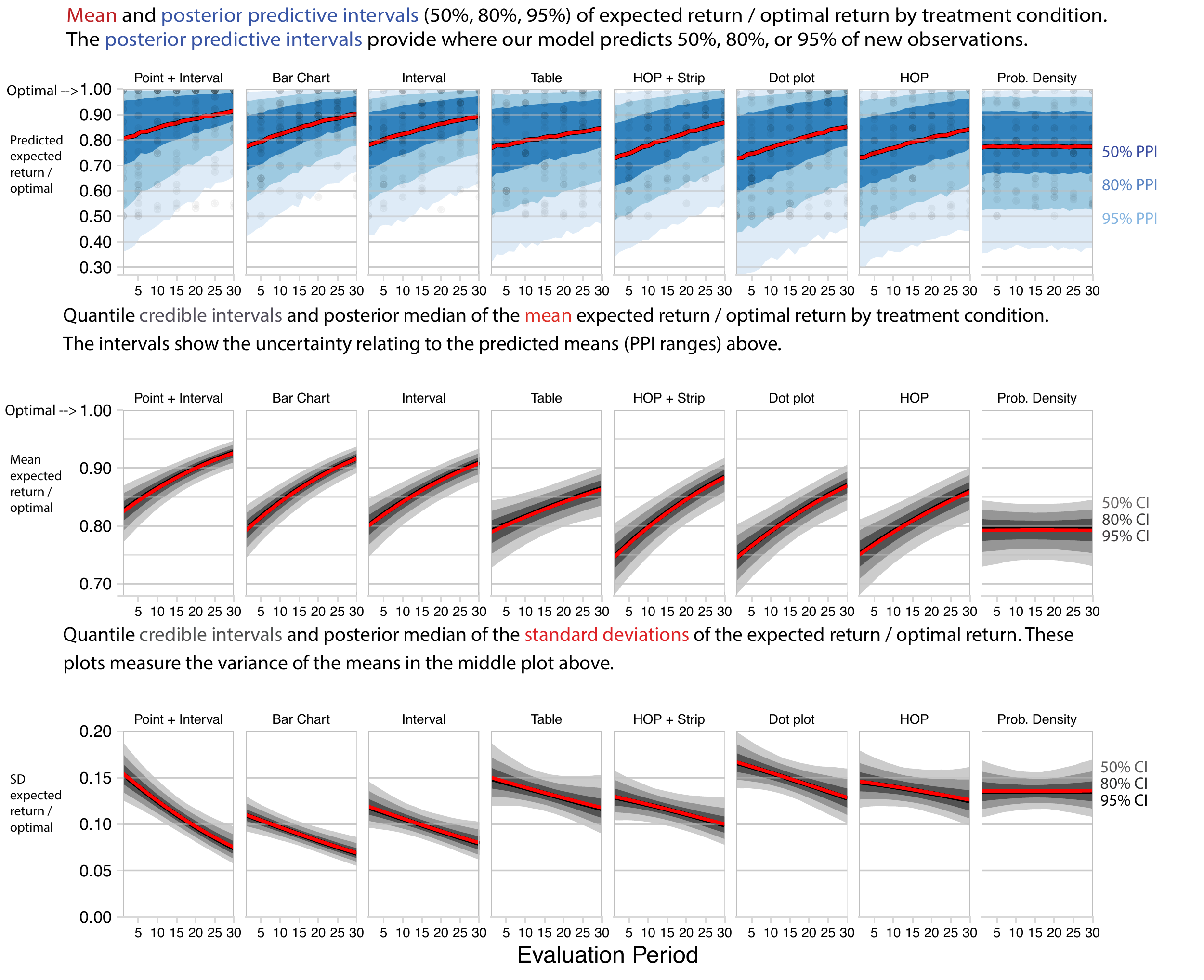}
\caption{Round 2 predicted mean, standard deviations, posterior predictive intervals, and credible intervals from model for expected return / optimal return by treatment and evaluation period. These figures are based on Fernandes \textit{et al.} \cite{fernandes2018uncertainty}.}
\label{fig:line-plot}
\vspace*{-3mm}
\end{figure*}

\section{Results}

\textbf{RQ1:} Consistent with Benartzi and Thaler \cite{benartzi1999risk}, we find evidence of myopic loss aversion with bar chart visualizations (round 1), as participants had a significantly lower stock allocation for 1 year versus 30 year evaluation period (Wilcoxon paired test on the median stock allocation, $V=2020.5$, $z = -9.188$, $p < .001$). 
Figure \ref{fig:boot-rq1} provides the bootstrapped 95\% confidence intervals ($n = 198$) for mean stock allocation. 
Notably, stock allocations for 1-year and 30-year evaluation periods were similar to the original study by Benartzi and Thaler \cite{benartzi1995myopic}, adding support for the robustness of the effect in our task and with a crowdsourced participant pool.

We also find a positive monotonic relationship between evaluation period and stock allocation (and expected return) (Figure \ref{fig:bootstraped-values}).  We noticed that the most significant allocation is the 1 year evaluation period in which average stock allocation was around 40-60\% (or 0.7-0.8 expected ratio). This makes sense as the 1 year had the largest amount of negative returns for stocks. We also find that participants' stock allocation tends to level off around 15 year evaluation period, especially for round 1. Figure \ref{fig:bootstraped-values} shows the monotonic relationship between evaluation period and participant's expected return normalized by optimal return from both round 1 (bar chart only) and round 2 (visualization treatment). We grouped the plots based on the similarity of second round performance and alignment between round 1 and 2. For example, consider the top row (best performing) plots had similar results in round 1 and 2. The Point + Interval and Interval plots led to the best expected returns comparing to the other uncertainty visualizations. 
But in the next two rows we can see a decrease of means into round 2, indicating that these visualizations may have hurt performance and increased the equity premium. One interesting exception is the probability density treatment, in which (especially for round 2) stock allocation was nearly flat across all evaluation periods. We'll examine this more carefully in our RQ2 modeling and in our discussion. 

\textbf{RQ2:} Figure \ref{fig:eval_end_plots} provides posterior estimates of the effects of each visualization by the dependent variable and its conditional mean and standard deviation. To measure the effect of uncertainty visualizations on investment decisions, we provide each treatment with posterior samples of the mean for the 1 year (blue) and 30 years evaluation periods. These plots show the credible intervals of the conditional mean (mu) and standard deviations for each condition. As noted by Fernandes \textit{et al.} \cite{fernandes2018uncertainty}, since these plots are conditional values, they show the mean and standard deviation for a typical participant given their treatment and evaluation period. 

First, we find evidence of differences in performance (equity premium) by visualization. Participants with the point + interval, bar chart, and interval visualizations had on average higher stock allocations and associated higher expected rates of return. Using that expected rate of return for the hypothetical retirement investor (37 year olds), we can estimate those allocations would lead to over \$600,000 value at retirement. Compare this example with participants who used the probability density. Consistent with results in \textbf{RQ1}, the model predicts decisions for the probability density with little variation. Those participants choose 60\% stocks, which on average would have led to around \$450,000 investment at retirement, a nearly two-thirds less value at retirement for 30 year evaluation period. 

Nevertheless, on average the decisions were still distant from the optimal (dotted line) as most participants choose the majority stocks but with some bonds. One interpretation of this distance is the equity premium, or how much investors would be willing to forgo to not fully invest in stocks (equity). Similarly, we can also measure myopic loss aversion, which would be the difference between the 1 year (blue) and 30 year evaluation periods (orange). We find that myopic loss aversion can vary between 20\% drop in stock for most visualizations. However, for the table and especially the probability density, we find less evidence of myopic loss aversion. 

Figure \ref{fig:line-plot} provides the model's \textit{prediction} of participant's decisions for their expected returns relative to the optimal expected return (i.e., 100\% stock allocation) as well as converted values. Similar to Fernandes \textit{et al.} \cite{fernandes2018uncertainty}, we use a Bayesian framework to enable measurements of the marginal posterior predictions to predict how a random participant would perform in our experiment's task. The top of Figure \ref{fig:line-plot} provides the mean expected return / optimal return (red line). Similar to RQ1, we find that on average the mean performance increases with the evaluation period (x axis) as on average participants allocate more stock when viewing returns in a longer evaluation period. One notable exception is the density plot which exhibits a flat mean performance, which is consistent with what we observed in Figure \ref{fig:bootstraped-values}.

In addition to the mean values, the figure also includes the posterior predictive intervals, or PPIs, for each treatment with three different value ranges: 50\% (dark blue band), 80\% (light blue band), and 95\% (lightest blue band). Let's consider the Point + Interval treatment to interpret these values. This treatment exhibited on average the highest value relative to the optimal strategy in which about 50\% of decisions (dark blue) ranged from 70-90\% optimal (1 year evaluation period) to 87-97\% optimal (30 year evaluation period). What's interesting across all of these plots is that while optimal (1.00) was possible within 95\% PPI values (lightest blue), for most treatments at best the optimal value was only within the 80\% PPI range, especially for longer evaluation periods. While myopic loss aversion and the visualization treatment accounts for some of the underallocation to stocks, this result indicates an additional missing factor that limits participants from making the optimal decision of 100\% stocks.

\section{Visual Reasoning Strategies}

To understand participants' reasoning in the task, we analyzed participants' qualitative feedback on the strategies they used to arrive at their decisions. These self-reported descriptions were recorded after each round. To expedite the the analysis, we used Non-Negative Matrix Factorization (NMF) topic modeling \cite{arora2012learning} to categorize comments from each round into several topics. Two researchers then qualitatively evaluated the semantic differences between the topics by reading top representative documents from each topic. This process resulted in several themes that summarize participants' strategies for each round.  

In round 1, participants experienced the bar chart visualization of expected percentage returns based on different evaluation periods. Based on 10 extracted topics, we observed three prevailing themes for round 1. The majority of the comments are related to users' strategies to allocate funds based on their perceptions of risk, returns, and balancing trade-offs between the two.

\textbf{Minimize risk or maximize reward:} A group of comments primarily focus on minimizing risk / losses. For example, one participant wrote: \textit{"I looked at the risk and amount of possible losses to make my decision."}. Another participant wrote: \textit{'I based it on risk. If a fund had a high likelihood of a negative return, I allocated less money to that fund.'}. One participant explicitly mentioned that although they look at highest returns, they primarily focus on minimizing losses: \textit{"I looked to see which had the best return. I also looked at negative returns. I tended to stay with the option that had little or no negative returns."}. Another group of comments are related to the strategy of maximizing returns. For example, a participant mentioned \textit{"I used the charts to determine what my best chances of the highest return would be."} Similarly, another participant said: \textit{"I looked out for which investment on the chart will give the highest return and allocated a larger percentage of my money to it".}

\textbf{Balancing gains and losses:} A large portion of the comments  described attempts to balance gains versus losses. For example: \textit{"I looked for which charts showed the biggest returns with the least risks of negative returns."} Another participant mentioned how they used the safer fund to hedge their losses: \textit{"[...] I tried to balance reward vs risk and allocate accordingly. I tended to invest more with the aggressive fund in order to maximize profits, but still invested some with the other fund to hedge my gamble."}

\textbf{Negative bars as a key decision aspect:} Other participants looked specifically for negative returns and preferred funds without them (e.g., evaluation periods over 15 years) to maximize returns. For example, one participant said: \textit{"If there were no negative years in either chart, I chose the chart that had the higher returns. If there were negative years in one chart that also had higher returns in other years, I chose 80\% or 90\% of the riskier chart, with 10\% or 20\% of the safer chart to balance out the bad years."} Another user had a similar strategy: \textit{"I saw whether there are negative bars. If yes, how many and how long. I tried to imagine the average of an all positive chart and compare that average with other chart."}

In round 2, many responses were consistent with round 1. This indicates some users replicated their strategies in both rounds. For example, one user within the density treatment mentioned that \textit{"I looked to see how much of the distribution was below zero. I viewed the larger area of distribution below zero as being more risky, so I invested more conservatively in those funds.  I attempted to invest more heavily in the less risky funds so that I would have a greater chance of not losing money."}. Another user in the table condition had a similar strategy: \textit{"I tended to go with slow growth again. Better if it didn't start out in the negatives cause it could quickly go back in the red zone."}

However, since users were randomly assigned to different visualization treatments, some topics naturally emerged as being related to each visualization treatment. Here we will describe some interesting emerging themes about different visualization techniques.

\textbf{Point plot and average rate of return:} Some users in the point treatment reported allocating more funds to higher (point) mean returns. For example, one user mentioned: \textit{"I looked at the average return dot. I put all money into the fund with the higher return."} Another mentioned that \textit{"I always chose the fund with the higher average return."} 

\textbf{Interval plot and confidence intervals:} Within the interval treatment, users explained how the dark green area with 66\% interval had a much larger influence on their decision, potentially limiting the effects of extreme values in datasets shown to users. For example, one user simply said: \textit{"Wanted the dark green to be highest."}. Similarly, another said: \textit{"I tried to see which chart had the larger opportunity for growth in the dark green zone.  I tried to focus more on the dark green zone than the light green zone."}

\textbf{HOPs and volatility:} Within these treatments, some users mentioned they allocated funds with samples with higher returns. For example, one user mentioned: \textit{"I tried to notice if one was higher than the other for most years and allocated more for that one."}. However, some users focused on volatility and made decisions based on funds that changed less. For example, a user mentioned that: \textit{"It was difficult to see the distribution. So I allocated everything to the one that changed the least."} Another user discussed similar strategies: \textit{"I looked at the fluctuations over the period of time and used that to determine the volatility of each investment. I chose the one that I felt was safest to protect my investment but I did always put some in both funds."}

\textbf{Table, max-min heuristic:} Some users reported using a heuristic to find the maximum and minimum returns, especially in the Table treatment. One participant wrote: \textit{"I tried to look for the highest and lowest rates in the chart. I struggled to get a good mental picture of each fund, so I tended to balance a little more.}. Similarly, another user said: \textit{"I checked the highest and lowest return value of each chart and compared them to make decision. Middle ones I ignored."}



\section{Discussion and Limitations}

Participants' performance in the investment task clustered by visualizations with similar encodings. The bar chart and interval plots (including with point) had the highest while HOPs and dot plot had on average lower values. Probability density and table had low myopic loss aversion. But if we consider these plots' performance in round 1 to round 2 from Figure 7, participants decreased their stock allocation indicating that these plots may have increased aversion towards stocks towards more stable bonds. 

\subsection{Bar-encoded plots had higher stock allocation} 

The point + interval, bar chart, and interval plots had the highest stock allocations especially for longer evaluation periods (80\%+). From Figure \ref{fig:eval_end_plots}, for the 30 year evaluation period those conditions also had the lowest standard deviations, which indicate participant consistency near the mean. Consistent with \cite{benartzi1999risk}, shorter evaluation periods like 1 year decreased stock allocation near to 60-65\%.
However, for these plots we also find slightly less myopic loss aversion as seen in the difference between the 1 year and 30 year allocations (about 20\% less allocation or about 1\% less returns) compared to other conditions. 
To put it into perspective, Figure \ref{fig:eval_end_plots} shows that such a difference in allocations could amount to approximately \$150,000 at retirement for an average 37 year old investor. 
We suspect that participants were able to get better mental models of the averages through either direct encoding (point + interval) or easy-to-calculate heuristics like mid-range of intervals or identifying the median bar in the bar charts.

 


\subsection{HOPs and dot plots may amplify risk aversion} 
We find that HOPs and dot plots had lower mean stock allocation (and thus lower expected returns) when comparing both within (round 1 bar chart, Fig. \ref{fig:bootstraped-values}) and between subjects (round 2 bar chart, Fig. \ref{fig:eval_end_plots}). We suspect that participants found it more difficult to estimate means or medians visually. For animated plots like HOPs, we found in the feedback participants who overly focused on volatility and may have struggled to identify the mean. However, we think performance with these plots is likely to be context dependent. For example, one simple remedy for these plots would be to overlay mean or medians. We believe with a mean, anchor HOPs and dot plot participants would perform better when optimizing expected returns. Alternatively, different investment objectives may yield more promise for these techniques (e.g., minimum return targets like 5\% which may place greater importance on tail risks). 

 

\subsection{Density and table have mixed results}


Similar to HOPs and dot plots, we found that the table and the density plots had lower mean stock allocation than bar-encoded plots. However, table and the density plot differ by displaying little myopic loss aversion as participants didn't change their stock allocation much with longer evaluation periods (see Figure \ref{fig:eval_end_plots}). But within these two plots, we suspect that participants had trouble differentiating either due to cognitive constraints (table) or continuity smoothing (density).

We find stock allocation (and expected returns) for the table was near the middle but showed some but not significant myopic loss aversion. Unlike HOPs and dot plots, table participants could approximate the means with heuristics like $(min + max) / 2$, which could explain the table's higher stock allocation.
However, the table group did worse than the bar-encoded plots perhaps because their heuristics weren't as accurate as estimates from bar-encoded plots as the average calculation was challenging.
We suspect some may have encoded the negative numbers more easily than other plots which could increase loss aversion. Alternatively, loss aversion may be reduced as the table made it hard to visually measure magnitude (i.e., how large are the losses).

Contrary to the other plots, there was no evidence that probability density plots led to myopic loss aversion.
This is despite finding that the same participants exhibited consistent myopic loss aversion in round 1 when presented with bar charts (50\% in 1 year and 80\% in 30 year, see Figure \ref{fig:bootstraped-values}). 
This is compared to a zero-to-negative difference for the same periods in round 2 with the density plot. 
We suspect that participants may have had difficulty in measuring central tendencies given density's smooth distribution, especially for longer evaluation periods. Like HOPs and dot plot, we suspect overlaying means may increase stock allocation.



\subsection{Limitations and Future Work}

This study focuses on long term retirement investment allocation and makes several assumptions that may limit the results. First, participants' incentives are based on expected results over a thirty year planning horizon. 
In practice, many retirement investors may want to plan for a near-term withdrawal (e.g., sell stocks for a major purchase or emergency). Second, like \cite{benartzi1999risk}, we consider only two funds that represent general asset classes (stocks and bonds). This ignores other asset classes like cash, real estate, or riskier assets (e.g., individual stocks, (crypto)currencies, high yield bonds). Third, we assume no effects from taxes or inflation (e.g., zero discount rate), which may affect actual investment behaviors for retirement. Fourth, we use a historical, non-parametric simulation (bootstrap with replacement) of past returns for stimuli and incentives. Instead, parametric approaches like monte carlo simulation based on continuous distributions (e.g., Normal, t-distribution, or fat tail distributions) would produce more uncertain returns and may benefit from visualizations with continuous representations of uncertainty \cite{kay2016ish,fernandes2018uncertainty}.

There are multiple avenues of potential future work. First, the goal of maximizing the expected returns is sensible but in practice retirement investors may have a slightly different decision problem. For example, many retirement investments are made to meet a balance goal, not necessarily to maximize expected return \cite{gunaratne2015informing}. Future work could modify the objective and incentivize reaching a simulated goal (e.g., reach 5\% annualized returns). 
Although bar-encoded representations of the mean or median appeared to aid performance in our task, different visual encodings which bring attention to the variability in returns or other statistics may be better suited to alternative objectives \cite{kale2020visual}.

Second, we did not provide users feedback on their decisions. Future work could explore learning effects through simulated investment feedback. 
Such a mechanism could measure ``explore-exploit'' trade-offs in allocation decisions. 
Recent research in cognitive science \cite{wulff2015short} has used a similar approach to examine short versus long-run strategies which either exploit known information (i.e., experienced returns) or explore different allocation strategies. 
Another possible direction is the interaction of uncertainty representations with descriptive text that on textual uncertainty \cite{hullman2018pursuit} or strategy cues \cite{wesslen2019investigating} to aid users how to interpret or interact with uncertainty representations \cite{kale2020visual,padilla2020uncertain}. 

Last, future work could expand experiment complexity by controlling for financial literacy, interaction with nudge-based heuristics common in day-to-day investment strategies \cite{benartzi2007heuristics}, or expand the study to experts like portfolio managers with shorter investment horizons, more complex assets, or more complex quantitative techniques. For example, we could incorporate more advanced financial risk measurements like VaR (value-at-risk) or Conditional VaR \cite{jorion2007value} and measure the interaction providing such metrics can modify the effects of uncertainty representations. Alternatively, we could expand to allocation across many assets and correlation of basket of n-assets, which could also be combined with Bayesian approaches in investment management \cite{kolm2021black} like Black-Litterman model that incorporate a financial manager's beliefs into a Markowitz modern portfolio theory (MPT) framework \cite{fabozzi2002legacy}.

\section{Conclusion}
In this paper, our contributions include findings from a crowdsourced (MTurk) incentivized mixed design experiment on the effect of uncertainty visualizations have on myopic loss aversion and the equity premium in long-term (retirement) investment decisions. Our results suggest that visualizations could have a large effect in \$100,000's balances at retirement for a typical long term investor. While myopic loss aversion has some variation across visualizations, performance remains sub-optimal (below 100\% stocks) which suggest future work in providing feedback and learning effects.

\acknowledgments{
The authors wish to thank the authors of Fernandes \textit{et al.} \cite{fernandes2018uncertainty} whose helpful  supplemental materials were heavily used for modeling.}

\bibliographystyle{abbrv-doi}

\bibliography{template}
\end{document}